# Confidence intervals for adaptive trial designs I: A methodological review

**David S. Robertson[1*], Thomas Burnett[2], Babak Choodari-Oskooei[3], Munya Dimairo[4], Michael Grayling[5], Philip Pallmann[6], Thomas Jaki[1,7]**

[1] *MRC Biostatistics Unit, University of Cambridge, UK*
[2] *University of Bath, UK*
[3] *MRC Clinical Trials Unit at UCL, UK*
[4] *School of Health and Related Research (ScHARR), University of Sheffield, UK*
[5] *Johnson & Johnson Innovative Medicine*
[6] *Centre for Trials Research, Cardiff University, UK*
[7] *University of Regensburg, Germany*
[*] *Corresponding author: david.robertson@mrc-bsu.cam.ac.uk*

## Abstract

Regulatory guidance notes the need for caution in the interpretation of confidence intervals (CIs) constructed during and after an adaptive clinical trial. Conventional CIs of the treatment effects are prone to undercoverage (as well as other undesirable properties) in many adaptive designs, because they do not take into account the potential and realised trial adaptations. This paper is the first in a two-part series that explores CIs for adaptive trials. It provides a comprehensive review of the methods to construct CIs for adaptive designs, while the second paper illustrates how to implement these in practice and proposes a set of guidelines for trial statisticians. We describe several classes of techniques for constructing CIs for adaptive clinical trials, before providing a systematic literature review of available methods, classified by the type of adaptive design. As part of this, we assess, through a proposed traffic light system, which of several desirable features of CIs (such as achieving nominal coverage and consistency with the hypothesis test decision) each of these methods holds.

## 1. Introduction

An adaptive design (AD) is a clinical trial design that allows for prospectively planned modifications to one or more aspects of the trial based on accumulating data from participants in the trial[1–3]. These planned modifications vary widely in their intent and scope, but carry a high-level commonality of increasing flexibility and improving efficiency, while maintaining trial integrity and validity. The most common types of AD include those that can select a patient (sub)population (adaptive enrichment designs), modify the randomisation ratio to, e.g., favour better performing arms (response-adaptive randomisation designs), revise the recruitment target based on an updated power calculation (sample size reestimation designs), select promising treatment(s) out of several experimental options (multi-arm multi-stage (MAMS) designs) and terminate the trial early for efficacy or futility (group sequential designs). For a more detailed overview, see, e.g., Bretz et al. (2009)[4], Pallmann et al. (2018)[1], Burnett et al. (2020)[5], and the PANDA online resource[6] (https://panda.shef.ac.uk/).

Whilst there is now a very large body of literature relating to ADs, the majority has focused on the key question of hypothesis testing, i.e., how to enable the inclusion of various types of adaptation while maintaining control of certain decision error rates. By comparison, computation of inferential quantities, such as treatment effect estimates and confidence intervals (CIs), has received comparatively less attention. Consequently, to stimulate the field and simultaneously offer practical assistance to researchers, a recent two-part series of articles[7,8] sought to (a) review available methods to remove or reduce bias in point estimates following an adaptive trial, (b) explore how to choose and implement an estimator, and (c) guide on how to report estimated effects.

Both of these articles on point estimation acknowledged the importance of methods for obtaining CIs following a trial utilising an AD, but left an in-depth discussion of this topic as out-of-scope. Indeed, whilst point estimates (summary measures of treatment effects) are often the primary focus of a study's final analysis as a core attribute of an estimand[9], capturing uncertainty around such estimates correctly is also essential to aid interpretation. It is CIs that capture this uncertainty by offering an interval that is expected to typically contain the unknown parameter of interest. When choosing the CI to calculate, an important consideration in practice is the desirable properties of a CI procedure. Principally, this relates to the CI having the desired coverage probability (i.e., the long-run probability that the CI contains the true unknown treatment effect of interest). However, it may also include numerous other considerations, including that the CI will indeed be an interval (i.e., it is not disjoint), that narrower CIs are preferred as they are more informative (reflecting less uncertainty), that the CI will always contain an associated point estimate, that it will always be consistent with the decision rule (i.e., with an associated hypothesis test), and that it is computationally feasible.

Put simply, the problem therein for ADs is then that use of standard CI methodology (i.e., CIs constructed using methods that do not account for the fact an AD has been used) for parameters of interest may not result in these desirable properties. Indeed, recent regulatory guidance highlights the important role CIs play, along with the pitfalls of utilising CI methods developed for conventional fixed-sample designs for ADs. Specifically, the U.S. Food and Drug Administration (FDA) notes that "confidence intervals for the primary and secondary endpoints may not have correct coverage probabilities for the true treatment effects" and thus "confidence intervals should be presented with appropriate cautions regarding their interpretation"[10]. It also notes the need to pre-specify methods used to compute CIs at the end of an adaptive trial (as also reflected in the ACE guidance[2,3]). The European Medicines Agency (EMEA) guidance takes an arguably stronger viewpoint, by stating "methods to … provide confidence intervals with pre-specified coverage probability are required" if an AD is going to be deployed in a regulated setting[11].

Thus the availability of methodology for CI construction, specific to ADs, is of critical importance. This has facilitated, for certain ADs, comparisons of available CI methods. See, e.g., comparisons for Simon two-stage trials[12], phase II/III trials[13], and (adaptive) group sequential trials[14,15]. However, a wider and more up-to-date overview of available methodology is needed to facilitate the use of these methods and to identify key open questions for future research. It is this overview we seek to provide here.

This article proceeds as follows. First, in Section 2 we elaborate on the issue of using 'standard' CI methodology for trials using an AD, discussing the potential problems this can cause, before describing several broad classes of available techniques for computing adjusted CIs suited to ADs in Section 3. Following this, in Section 4 we provide a literature review of available methods for constructing CIs for ADs, including a traffic light system that categorises which of several desirable features each method holds. We then conclude in Section 5 with a discussion of the current landscape of methods for computing CIs after an adaptive trial.

## 2. Potential problems with using standard CIs for ADs

In a traditional (fixed) design for a clinical trial with a single primary outcome, the sample space of that outcome is one-dimensional and so the construction of CIs (just like p-values) at the end of the trial is relatively straightforward (at least for commonly encountered continuous outcomes; for discrete outcomes there is additional complexity). This is often achieved by 'inverting' the hypothesis test to obtain desired CI bounds around the maximum likelihood estimate (MLE), as demonstrated more explicitly in Section 3. These 'standard' CIs often produce desired coverage under correct distributional assumptions of the parameter of interest and are consistent with the test decision.

On the contrary, in ADs, several factors may render the use of standard CIs inappropriate and complicate the methods for deriving appropriate CIs for ADs at the end of the trial. First, the possible outcomes at the end of the trial depend on the timing of interim analyses, observed results, and adaptation rule considered. As such, the outcome sample space is no longer one-dimensional but multi-dimensional, as it now includes the stopping stage of the trial (for example).

Second, standard CIs are often constructed based on distributional assumptions that are no longer met in an AD. For example, a selection, enrichment or stopping rule in an AD may mean that the distribution of a parameter of interest is no longer normal, but the standard CI assumes an underlying normal distribution. This can then lead to a disconnect between the test decision (which does account for the adaptive features of the design such as truncation of the parameter distribution due to selection) and the standard CI. As a result, a test decision and the decision derived from the standard CI (i.e., whether it includes the null effect) may differ, which leads to substantial problems in interpretation and communication of the trial results, see the real data example for an adaptive enrichment design in Wassmer and Draglin (2015)[16].

Third, as reflected by Robertson et al. (2023a,b)[7,8], the MLE after an AD is potentially biased in the sense that it will tend to systematically deviate from the true value; and standard CIs tend to be centred around this biased estimate. Finally, the use of standard CIs that do not account for the adaptive nature of the design tend to produce incorrect coverage, often lower than the desired nominal coverage, although higher coverage may also occur. Just like the statistical bias of standard point estimates, the level of incorrect coverage of standard intervals can be impacted by many factors including the underlying treatment effect, trial adaptation considered, decision rules (e.g., stopping boundaries), or the probability of triggering adaptations (e.g., stopping or selection). This incorrect coverage of standard CIs, which also depends on the magnitude of the treatment effect, may lead to challenges around the overall interpretation of the study results, as well as their use in secondary research such as meta-analyses, systematic reviews, and even health economic evaluations.

# 3. Constructing CIs for ADs

## 3.1 Fundamental concepts and definitions

We first introduce some fundamental concepts and definitions for CIs in general. Suppose we have a random sample $X$ from a probability distribution with parameter $\theta$, which is the single parameter of interest in the trial (we defer the case for multiple parameters of interest until Section 4.1). A CI for $\theta$ with confidence level (or confidence coefficient) $1 - \alpha$ is a random interval $(L(X), U(X))$ that has the following claimed property: $P(L(X) < \theta < U(X)) = 1 - \alpha$ for all $\theta$. (Note that sometimes this is replaced by $P(L(X) < \theta < U(X)) \geq 1 - \alpha$ in the literature, particularly for discrete outcomes where it may not be possible to achieve equality for all values of $\theta$).

The coverage probability (often shortened to just 'coverage') of a CI is given by $P(L(X) < \theta < U(X))$. The confidence level in the definition above is the 'nominal' coverage probability. If all assumptions used in deriving a CI are met, this nominal coverage probability will equal the (actual/true) coverage probability. However, if these assumptions are not met, such as in the context of many ADs, then the actual coverage may be greater than the nominal coverage probability (known as overcoverage, and the CI is termed a conservative CI) or less than the nominal coverage probability (known as undercoverage, and the CI is termed an anti-conservative CI). Arguably, in such situations the CI may be more accurately described simply as an 'interval' since it does not have the desired confidence level.

Another important concept for CIs is the distinction between *one-sided* vs *two-sided* CIs, which correspond (at least in theory) with one-sided vs two-sided hypothesis testing. To fix ideas, consider the case where $\theta$ takes values on the real line, i.e. $\theta \in (-\infty, +\infty)$. A two-sided CI corresponds with a two-sided hypothesis test of $H_0 : \theta = \theta_0$ versus the alternative $H_1 : \theta \neq \theta_0$, and would be of the form $(l, u)$ where $l$ and $u$ both take finite values. In contrast, a one-sided CI corresponds to a one-sided hypothesis test of $H_0 : \theta = \theta_0$ versus the alternative $H_1 : \theta > \theta_0$ (or $H_1 : \theta < \theta_0$), and would be of the form $(l, \infty)$ (or $(-\infty, u)$). In practice, for one-sided hypothesis tests it is common to replace a $1 - \alpha$ level one-sided CI $(l, \infty)$ with a $1 - 2\alpha$ level two-sided CI $(l', u)$ with $l' = l$ and $u < \infty$, or a $1 - \alpha$ level one-sided CI $(-\infty, u)$ with a $1 - 2\alpha$ level two-sided CI $(l, u')$ with $u' = u$ and $l > -\infty$.

## 3.2 Inverting a hypothesis test statistic

A widely used technique for constructing CIs is by exploiting the duality between hypothesis tests and CIs. Suppose we are testing a null hypothesis $H_0$ for a parameter of interest $\theta$ with a corresponding level-$\alpha$ hypothesis test procedure. If $H_0$ is true, a $100(1 - \alpha)\%$ CI corresponds to the values of $\theta$ for which $H_0$ is not rejected at level $\alpha$. This implies[17] that rejecting $H_0$ whenever such a $100(1 - \alpha)\%$ CI does not include the null effect is equivalent to rejecting $H_0$ whenever the level-$\alpha$ test procedure yields a $p$-value of less than $\alpha$. This guarantees that the CI is *consistent* with the hypothesis testing decision (see Section 3.4).

Of note, whilst the $100(1 - \alpha)\%$ CI is the set of values of $\theta$ for which $H_0$ would not be rejected, the acceptance region of a level-$\alpha$ test is the set of values of the test statistic for which $H_0$ would not be rejected. An inversion of the test procedure is achieved by 'mapping' these two sets of values onto each other, thereby using the acceptance region of an existing hypothesis test to identify a set of values for the parameter of interest (i.e., a confidence set) which is consistent with not rejecting $H_0$. In practice, one would typically 'invert' the equation for the test statistic corresponding to the boundary values, thereby identifying the CI bounds. Where this cannot be done analytically, numerical approaches can be employed to find the CI bounds.

A key example in the context of ADs are repeated CIs (RCIs) used for group sequential designs (as well as MAMS designs). We defer the motivation and general definition of RCIs to Section 4.2.1, and only describe their construction for group sequential designs in terms of inverting a hypothesis test statistic as described in Jennison and Turnbull (1999)[18]. Consider a two-sided group sequential test of the hypothesis $H_0 : \theta = \theta_0$ with type I error probability $\alpha$. This has the form:

$$\text{Reject } H_0 \text{ at stage } k \text{ if } |Z_k(\theta_0)| \geq c_k(\alpha), k = 1, \ldots, K$$

where $Z_1, \ldots, Z_K$ are the cumulative standardised test statistics and $c_1(\alpha), \ldots, c_K(\alpha)$ are the group sequential critical values. The RCI at stage $k$, denoted $I_k$, is defined by inverting this group sequential test, i.e., by defining $I_k = \{\theta_0 : |Z_k(\theta_0)| < c_k(\alpha)\}$.

Another class of CIs for group sequential designs that are also based on inverting a hypothesis test statistic are 'exact' CIs (see also Section 4.2.1). However, rather than working with the group sequential test as described above, exact CIs rely on a chosen ordering of the sample space to determine which values of the pair $(k_T, Z_T)$ are more extreme evidence against the null hypothesis, where $k_T$ denotes the stage the trial stops at and $Z_T$ denotes the observed cumulative standardised test statistic. Given the chosen ordering, one can derive an acceptance region and corresponding hypothesis test of $H_0: \theta = \theta_0$ with type I error probability $\alpha$. These hypothesis tests can then be inverted to give a CI for $\theta$. Depending on the ordering chosen, the resulting CI may not necessarily agree with the original group sequential test above; for further discussion see Jennison and Turnbull (2000)[18(chap8.5)].

## 3.3 Constructing a pivotal quantity

As an alternative to inverting the distribution one may instead construct CIs through using a pivotal quantity[17]. Suppose one has some data $\boldsymbol{X} = (X_1, \ldots, X_n)$ and an (unknown) parameter of interest, say $\theta$ (note this may be a vector of parameters but for convenience we will consider only a single parameter). Then a pivotal quantity is defined as a function of the data $X$ and the parameter $\theta$, denote this by $g(X, \theta)$, where the distribution of $g(X, \theta)$ does not depend on $\theta$.

Assuming the distribution of $g(X, \theta)$ is known one may find $l$ and $u$ such that
$$P(l \leq g(X, \theta) \leq u) = 1 - \alpha.$$
Thus with some manipulation one may find corresponding $\hat{\theta}_l$ and $\hat{\theta}_u$ such that
$$P(\hat{\theta}_l \leq \theta \leq \hat{\theta}_u) = 1 - \alpha,$$
and thus a $100(1 - \alpha)\%$ CI for $\theta$ is given by
$$[\hat{\theta}_l, \hat{\theta}_u].$$

To solidify this concept let us consider an example using the normal distribution. Suppose we have independent normally distributed data $\boldsymbol{X} = (X_1, \ldots, X_n)$ with an unknown mean $\theta$ and known variance $\sigma^2$, that is
$$X_i \sim N(\theta, \sigma^2) \text{ for all } i = 1, \ldots, n$$
To construct a $100(1 - \alpha)\%$ CI for $\theta$ we shall use a pivotal quantity. Let $\bar{X} = \frac{1}{n} \sum_{i=1}^{n} X_i$
then the pivotal quantity is given by
$$g(X, \theta) = \frac{\bar{X} - \theta}{\sigma / \sqrt{n}}$$
where conveniently we thus know that $g(X, \theta) \sim N(0,1)$. Defining $\Phi^{-1}(\ldots)$ to be the inverse cumulative distribution function (CDF) of the standard normal distribution we may then find a CI for the pivotal quantity by choosing $l = -\Phi^{-1}(1 - \alpha/2)$ and $u = \Phi^{-1}(1 - \alpha/2)$ satisfying the condition that
$$P(l \leq g(X, \theta) \leq u) = 1 - \alpha.$$
With a little work we can thus find the CI,
$$1 - \alpha = P(l \leq g(X, \theta) \leq u)$$
$$= P(\bar{X} - \Phi^{-1}(1 - \alpha/2) \, \sigma / \sqrt{n} \leq \theta \leq \bar{X} - \Phi^{-1}(1 - \alpha/2) \sigma / \sqrt{n})$$
Thus a $100(1 - \alpha)\%$ CI for $\theta$ is given by
$$[\bar{X} - \Phi^{-1}(1 - \alpha/2) \, \sigma / \sqrt{n}, \bar{X} - \Phi^{-1}(1 - \alpha/2) \sigma / \sqrt{n}].$$

It is useful to note the consequence of these well-known results in the context of the normal distribution. It is regularly the case that for the ADs that we consider for this work we wish to provide inference for the sample mean of our data in which case for sufficiently large $n$, via the central limit theorem, a normal approximation will often suffice. The extension of these methods to ADs depends on the situation in which they are applied but this core concept of finding a pivot corresponding to each estimate for which we require a CI may be applied. This has been applied in several works, see Section 4.4.

As an example of the application of finding an (approximate) pivot for ADs, Woodroofe (1992)[19] proposed the following in the context of group sequential designs with parameter of interest $\theta$. Let $Z_T$ and $I_T$ denote the cumulative standardised test statistic and (Fisher) information level, respectively, at the stage the trial stops (denoted $T$). If the sample sizes were fixed then the statistic $Z_T{}'(\theta) = \frac{Z_T - \theta I_T}{\sqrt{I_T}}$ would (asymptotically) follow a standard normal distribution. However, due to the potential for early stopping, this is not the case. Instead, the following modification of the statistic $Z_T{}'(\theta)$ gives a new statistic that more closely follows a standard normal distribution: $Z_T{}^*(\theta) = \frac{Z_T{}'(\theta) - \mu(\theta)}{\sigma(\theta)}$ where $\mu(\theta)$ is the mean of $Z_T{}'(\theta)$ and $\sigma(\theta)$ is the standard deviation of $Z_T{}'(\theta)$.

## 3.4 Bootstrap/resampling approaches

The approaches discussed so far in this Section require knowledge of a suitable distribution for the estimate, either directly or asymptotically. Bootstrap and other resampling based methods bypass the need for this by constructing approximate CIs through resampling from the data. To construct a CI for some estimate the bootstrap method will sample from the data (with replacement) repeatedly each giving an estimate of the parameter of interest. These estimates are used to approximate the distribution of the parameter from which we estimate the CI.

To give a little more insight let us consider our previous example from Section 3.3. We have data $\boldsymbol{X} = (X_1, \ldots, X_n)$ and wish to make inference on the mean $\theta$. We may estimate this by

$$\hat{\theta} = \frac{1}{n} \sum_{i=1}^{n} X_i$$

supposing this has a known distribution (or through use of the other methods in this Section) we might then construct an appropriate $100(1 - \alpha)\%$ CI.

Alternatively, without making any distributional assumptions we may construct this CI using bootstrap resampling[17]. We draw $M$ bootstrap samples of size $n$ from these data by sampling the data uniformly at random with replacement, denoting these samples by $\boldsymbol{X}^{(i)} = (X_1{}^{(i)}, \ldots, X_n{}^{(i)})$ for $i = 1, \ldots, M$. From each sample we construct the corresponding estimate $\hat{\theta}^{(i)} = \frac{1}{n} \sum_{i=1}^{n} X_j{}^{(i)}$.

For large $n$ the distribution of these $\hat{\theta}^{(i)}$ converges to the unknown distribution of $\hat{\theta}$. Without loss of generality we re-index these bootstrap estimates such that they are in order with

$$\hat{\theta}^{(1)} \leq \hat{\theta}^{(2)} \leq \ldots \leq \hat{\theta}^{(M)}.$$

We then have that

$$P(\hat{\theta}^{(M(\alpha/2))} \leq \theta \leq \hat{\theta}^{(M(1-\alpha/2))}) \approx 1 - \alpha$$

and thus an approximate $100(1-\alpha)\%$ CI for $\theta$ is given by

$$[\hat{\theta}^{(M(\alpha/2))}, \hat{\theta}^{(M(1-\alpha/2))}].$$

This is the so-called 'percentile interval' which is the simplest, but other (potentially better) choices of bootstrap CI exist[20].

This concept can be directly extended to the setting of ADs by using a bootstrap procedure as above while accounting for the adaptive nature of the design. For any given estimator, one may resample from the corresponding data used in constructing the estimate and use this in the construction of an appropriate CI. Such methods can be found in the methodological summary in Section 4.4.

As an example of the use of bootstrap for ADs, consider the context of response-adaptive randomisation for a multi-armed trial with $K$ arms and binary endpoints, where the allocation probabilities are adapted based on the accumulated patient response data. Rosenberger and Hu (1999)[21] describe how to use a bootstrap procedure in this context:

1) Obtain the observed data from the trial, i.e. the vector of observed success proportions $\hat{P} = (\hat{p}_1, \ldots, \hat{p}_K)$ and sample sizes on each arm $N = (n_1, \ldots, n_K)$.
2) Simulate the response-adaptive allocation rule $M$ times, using $\hat{p}$ as the assumed true response probabilities.
3) Compute bootstrap estimates of the vector of response probabilities $\widehat{P_1}^*, \ldots, \widehat{P_M}^*$ and sample sizes $N_1^*, \ldots, N_M^*$ from the simulations.
4) For each $i = 1, \ldots, K$ order $\hat{p}_i^{*1}, \ldots, \hat{p}_i^{*M}$ to form the ordered sequence $\hat{p}_i^{*(1)}, \ldots, \hat{p}_i^{*(M)}$
5) The simplest $100(1-\alpha)\%$ CI for $p_i$ is then ($\hat{p}_i^{*(M\alpha/2)}$, $\hat{p}_i^{*(M(1-\alpha)/2)}$).

## 3.5 Other approaches

### 3.5.1 Bayesian approaches

With the growing popularity of Bayesian methods for ADs (and clinical trials more generally) there is also growing interest in using the Bayesian paradigm for inference about the treatment effect. However, CIs are inherently a frequentist concept involving the repeated resampling or realisations of the adaptive trial in question. In contrast, from a Bayesian perspective the uncertainty around the parameter of interest, $\theta$, is quantified in terms of a credible interval, based on the posterior probability density of $\theta$ itself. Nonetheless, the frequentist properties of credible intervals (such as coverage) could in theory be investigated, although this was out of scope of our literature review and our paper more generally. A complicating factor is that the properties of credible intervals would additionally depend on the choice of prior distribution. An interesting approach that combines aspects of both Bayesian and frequentist methods is the confidence *distribution* approach as described by Marschner (2023)[22]. These confidence distributions provide a posterior-like probability distribution that does not require the specification of priors, and is compatible for frequentist inference.

### 3.5.2 Hybrid approaches

Finally, it is also possible to combine the approaches described above to construct CIs for ADs. For example, Chuang and Lai (1998, 2000)[23,24] proposed a hybrid technique consisting of elements of both exact and bootstrap methods to construct CIs and regions following (group) sequential trials. Specifically, they suggested using an exact method based on inverting a hypothesis test but replacing the normal quantiles with quantiles from a bootstrap distribution. Whilst this hybrid method does not have exact nominal coverage, it can be proven to yield second-order accurate CIs, and in simulations it has been shown to generally have close-to-nominal coverage and to outperform both exact and bootstrap methods in situations where these two have poor coverage.

A different kind of hybrid strategy was discussed by Kimani et al. (2014)[13] who compared different CI methods for seamless phase II/III trials in simulations and noted that the properties of CIs could be optimised by combining lower and upper bounds based on different methods, e.g., a lower bound based on a method known to have high power (to maximise the chance of rejecting the null hypothesis) with an upper bound based on a method which has exact or close-to-nominal coverage to provide an accurate upper limit for the size of the treatment effect. Such a compound CI would inevitably be asymmetric in most cases.

## 4. Methodological review of adjusted CIs for ADs

### 4.1    Search strategy and paper selection

We conducted a database search of Scopus on 21 August 2024 of all available papers (not including preprints) up to that date. We used a "title, abstract, keywords" search, with the following predefined search term: *(("confidence interval" OR "confidence region" OR "confidence limit" OR "confidence band") AND ("adaptive design" OR "adaptive trial" OR "adaptive clinical trial" OR "group sequential" OR "sequential trial" OR "sequential clinical trial" OR " drop the loser" OR "response adaptive" OR "sample size re-estimation" OR "seamless phase" OR "multi arm multi stage" OR "adaptive enrichment"))*.

Our search strategy retrieved a total of 324 papers, of which 179 were excluded as they were not relevant based on the title and abstract. We then looked for additional relevant papers citing or cited by the remaining 145, which added 27 papers. We conducted a full text review of these 172 papers. Information about the trial contexts, methodology used, advantages, limitations, code availability and case studies was extracted for qualitative synthesis. Full results giving a summary of each paper can be found in the Supporting Information.

Before presenting a summary of the results of the literature review in Section 4.4, we first briefly describe some key methodological concepts found in the literature on adjusted CIs for ADs in Section 4.2, followed by a discussion of desirable criteria for CIs in Section 4.3.

## 4.2    Key methodological concepts in the literature

### 4.2.1 Repeated vs final CIs

In an AD, CIs are used for end-of-study interpretation of results, i.e., at the final analysis of the trial data (hence the term 'final' CI). However, CIs can also be used for monitoring the trial and providing quantification of uncertainty around treatment effect estimates at *interim* analyses/looks at the accumulating data. One key example of this is when presenting data to a Data and Safety Monitoring Board (DSMB) for deciding whether or not to stop a trial early for efficacy or lack-of-benefit/futility.

This second motivation leads to the concept of *repeated* CIs (RCIs), as already briefly introduced in Section 3.2, which are a sequence of interval estimates for $\theta$ that can be calculated at *any* interim look at the trial data. More formally, given a trial with a maximum of $K$ stages (or equivalently $K - 1$ interim analyses) the RCIs for a parameter $\theta$ are defined[18] as a sequence of intervals $I_k$, $k = 1, ..., K$, for which simultaneous coverage probability is maintained at level $1 - \alpha$, so that the following equation holds:

$$Pr_\theta(\theta \in I_k \text{ for all } k = 1, ..., K) = 1 - \alpha \text{ for all } \theta.$$

Note that in fact if $\tau$ is any random stopping time for the trial (i.e., the stage the trial stops at, taking values in $\{1, ..., K\}$) then the above equation implies that $Pr_\theta(\theta \in I_\tau) = 1 - \alpha$ for all $\theta$. Hence, RCIs can be used while still maintaining the $1 - \alpha$ confidence limit *regardless of how the decision to stop the study was reached.*

Consequently, final CIs can only be calculated in a valid way at the stage a trial stops according to a pre-specified stopping rule, whereas RCIs are not tied to a stopping rule and can be calculated at any interim analysis. However, this means that RCIs will typically be wider than final CIs, particularly for earlier interim analyses[25]. Note that particularly in the group sequential literature the term 'exact' CI (as introduced in Section 3.2) often refers to final CIs rather than RCIs.

### 4.2.2   Individual CIs vs simultaneous confidence sets

Thus far, we have focused on the case where there is a single hypothesis and corresponding parameter $\theta$ of interest. However, many types of ADs consider multiple hypotheses and corresponding parameters simultaneously. MAMS designs are a key example, with multiple treatment arms being compared against a common control arm. To fix ideas, suppose we have null hypotheses $H_{01}, \ldots, H_{0J}$ with corresponding parameters of interest $\theta_1, \ldots, \theta_J$. Given a random data sample $X$, suppose we calculate individual two-sided CIs $(L_i(X), U_i(X))$ for each parameter independently. These will achieve the correct *individual* (also known as marginal or univariate) coverage, i.e., $P(L_i(X) \leq \theta_i \leq U_i(X)) = 1 - \alpha$ for $i = 1, \ldots, J$. However, the *overall* or *simultaneous* coverage probability of the $J$ CIs is not necessarily controlled, i.e., $P(L_1(X) \leq \theta_1 \leq U_1(X), \ldots, L_J(X) \leq \theta_J \leq U_J(X)) \neq 1 - \alpha$.

To achieve control of the simultaneous coverage probability requires specification of a multivariate *confidence set* or *confidence region* C(X), with the property that $P(\boldsymbol{\theta} \in C(X)) = 1 - \alpha$, where $\boldsymbol{\theta} = (\theta_1, \ldots, \theta_J)$. In general, such confidence sets are not necessarily a cross product of CIs. In order to recover CIs for each parameter, one can enlarge the confidence region to fit within a multidimensional rectangle. This comes with the disadvantage that the resulting CIs may be inconsistent with the test decision (see Section 4.3), as noted by Posch et al. (2005)[26].

In a trial that tests multiple hypotheses, when considering constructing CIs for each parameter of interest it is therefore important to decide whether correct individual coverage or simultaneous coverage is desired. This is closely linked to the concept of adjusting for multiplicity, i.e., whether one wants to control the marginal type I error rate (known as the pairwise type I error rate in the context of comparing with a common control arm) for each hypothesis at level $\alpha$ or controlling the overall familywise error rate (FWER), which is the probability of making at least one type I error. If there is interest in controlling the FWER, then the CIs should reflect this and hence simultaneous confidence sets/intervals should be considered. Conversely, if no adjustment for multiplicity is required (e.g., when the hypotheses tested are independent) then the usual individual CIs would be sufficient.

### 4.2.3 Conditional vs unconditional CIs

A final key distinction of CIs for ADs is whether they have the correct coverage *conditionally* or *unconditionally*. Intuitively, unconditional coverage refers to the coverage averaged across all possible realisations of an adaptive trial. In contrast, conditional coverage refers to the coverage averaged over a particular subset of trial realisations. For example, we might be interested in the coverage of the CI conditional on a trial continuing to the final stage, or a particular treatment arm being selected at the final analysis. More precisely, and returning to the setting with a single parameter of interest $\theta$, the conditional coverage of a CI $(L(X), U(X))$ is defined as $P(L(X) \leq \theta \leq U(X) \mid S)$, where $S$ is a particular (random) event of interest, such as the stopping stage of a group sequential trial.

A detailed discussion of the merits of conditional vs unconditional inference, including CIs, is beyond the scope of this paper. We refer the interested reader to Strickland and Casella (2003)[27], Fan and DeMets (2006)[28], Marschner and Schou (2019)[29] and Marschner et al.

(2022)[30] for useful discussion about conditional CIs in the context of group sequential trials. More general discussion and a proposed framework to view about conditional vs unconditional inference for ADs can be found in Marschner (2021)[31].

## 4.3    Desirable criteria for CIs

In addition to coverage (see Section 3.1), there have been other proposed desirable criteria for CIs in the literature. The main criteria include:

- Correct coverage (arguably essential)
- Width (all other things being equal, a smaller width is desirable)
- Consistency/compatibility with the hypothesis test (see below)
- Contains the point estimate of interest
- Is informative (see below)
- Is in fact an interval (i.e., not a union of disjoint intervals, or the empty set)
- Is computationally feasible/simple

A CI is *consistent/compatible* with the hypothesis testing decision if it excludes the parameter value(s) that are rejected by the hypothesis test, and conversely includes the parameter value(s) that are *not* rejected by the hypothesis test. If a CI is *not* consistent/compatible with the hypothesis testing decision then this can lead to problems with study interpretation and the communication of results. In theory at least, it is possible to 'invert' the hypothesis test used (see Section 3.2) to obtain a CI that is always consistent with the hypothesis test decision, but not all CIs are constructed in this way as seen in Section 3.

A CI is *informative* if it restricts the possible parameter space. For example, if $\theta$ corresponds to the success probability of a binomial distribution then a two-sided CI needs to be strictly contained in [0,1]. More formally, consider a parameter $\theta$ that takes values in the set $(a,b)$, where $a$ and $b$ may be infinite. Clearly, it is desirable that for a two-sided CI $(L(X), U(X))$, we have $L(X) > a$ and $U(X) < b$. Similarly, for an upper one-sided CI, it is desirable that $U(X) < b$, while for a lower one-sided CI, it is desirable that $L(X) > a$.

Note there is a stricter definition of a CI being informative for CIs that are compatible with a corresponding hypothesis test. Aside from the criteria above, a CI in this case is informative if it additionally provides more information than the hypothesis testing decision. For example, consider testing the null hypothesis $H_0: \theta = \theta_0$ vs the alternative $H_1: \theta > \theta_0$. If $H_0$ is rejected, then a CI is only informative if $L(X) > \theta_0$ .

## 4.4 Summary of literature review

As part of our summary of the literature review, we use a 'traffic light' system for the different classes of methods for constructing CIs, following the suggested desirable criteria for CIs given above. For simplicity, in the definitions that follow we consider the setting with a single parameter of interest $\theta$ and corresponding method for constructing a two-sided CI denoted $(L(X), U(X))$ given a random data sample $X$ with target coverage (i.e., claimed confidence level) of $1 - \alpha$.

*Coverage*

Green: CI has (actual) coverage equal to $1 - \alpha$ for all values of $\theta$, i.e., $P(L(X) \leq \theta \leq U(X)) = 1 - \alpha$ for all $\theta$. We add labels 'A' for Analytical and 'S' for Simulation if this property has been shown analytically or only by simulation, respectively.

Amber: CI has (actual) coverage greater than or equal to $1 - \alpha$ for all values of $\theta$, i.e., $P(L(X) \leq \theta \leq U(X)) \geq 1 - \alpha$ for all $\theta$. We add labels 'A' for Analytical and 'S' for Simulation as above.

Red: CI has (actual) coverage less than $1 - \alpha$ for at least one value of $\theta$, i.e., $P(L(X) \leq \theta \leq U(X)) < 1 - \alpha$ for some $\theta$.

*Interval*

Green: For all realisations of X, the method will result in a single CI (L(X), U(X)) for all $\alpha \in (0,1)$.

Red: For some realisations of X, the method does *not* result in a single CI (L(X), U(X)) for some $\alpha \in (0,1)$.

*Consistent*

Green: The CI is always consistent as it is constructed by inverting the hypothesis test used by the AD.

Red: The CI can be inconsistent i.e., there are explicit examples of where the CI and hypothesis test decision for the AD are conflicting.

*Informative*

Green: For all realisations of X, the procedure will result in an informative CI for all $\alpha \in (0,1)$.

**Red**: For some realisations of X, the procedure does *not* result in an informative CI for some $\alpha \in (0,1)$.

*Contains MLE*

In Section 4.3, one of the desirable criteria for CIs is that it "contains the point estimate of interest". A detailed discussion of what the point estimate of interest could be for an AD is out of scope of this paper, and we refer the reader to Robertson et al. (2023a,b)[8,32]. For the purposes of the literature review, we simply use the usual end-of-trial MLE as our point estimator of interest, both because this is the most commonly reported estimator and because this has been proposed as a criterion for assessing CIs[18]. Note that in theory, a CI could contain an unbiased or bias-reduced point estimate but not the MLE.

**Green**: For all realisations of X, the procedure will result in a CI that contains the MLE for all $\alpha \in (0,1)$.

**Red**: For some realisations of X, the procedure does *not* result in a CI that contains the MLE for some $\alpha \in (0,1)$.

*Computation*

We note that for calculating a single CI (i.e., for a single trial realisation), there are no computational concerns for almost all methods given modern computational power. However, when it comes to assessing the performance of CIs through simulation studies (e.g., at the planning stage of the trial), computation can still be a (major) limitation.

**Green**: CI can be calculated using analytical formulae that can easily be evaluated using standard statistical software or code (such as R).

**Amber**: Calculation of the CI involves the use of numerical optimisation and/or computer simulation.

Armed with this traffic light system, we summarise the results of our literature review in Table 1 by classifying the CI methods used for each broad class of AD. For each combination of design class and CI method, we provide a summary of any key features of the CI method reported in the literature, list (key) references in the literature (categorised by outcome type, for example) and also show how the CI method performs as assessed by the traffic light system.

| Design | Method(s) | Coverage | Interval | Consistent | Informative | Contains MLE | Computation |
|---|---|---|---|---|---|---|---|
| **Group sequential** | ***Exact confidence intervals***<br><br>*Tsiatis et al. (1984)[33], Jennison and Turnbull (1999)[18]*<br><br>• Colours on the right assume the use of stage-wise ordering of the sample space and that the canonical joint distribution of the group sequential test statistics holds exactly (e.g., for normally distributed data with a known variance - see Jennison and Turnbull, 1999[18])<br><br>• For an example of where the CI does not contain the MLE, see Tsiatis et al. (1984)[33]<br><br>• Can fail to be an interval if other orderings of the sample space are used, see e.g., Rosner and Tsiatis (1988)[14], Emerson and Fleming (1990)[34]<br><br>• Coverage can be conservative when exact method is used for binary outcomes, and there can be inconsistency with the test decision; see Lloyd (2021)[35] for a discussion of these features<br><br>• For non-normal outcomes, a 'hybrid' strategy can be used which combines the exact CI approach with resampling, see Chuang and Lai (1998, 2000)[23,24], Lai and Li (2006)[36]<br><br>• Conditional perspective considered in Strickland and Casella (2003)[27], Ohman (1996)[37], Fan and DeMets (2006)[28], Koopmeiners et al. (2012)[38], Marschner and Schou (2019)[29], Marschner et al. (2022)[30]<br><br>**See also**<br>*Siegmund (1978)[39], Kim and DeMets (1987)[40], Rosner and Tsiatis (1988)[14], Chang (1989)[41], Facey and Whitehead (1990)[42], Emerson and Fleming (1990)[34], Wittes (2012)[43], Hampson et al. (2017)[44], Hanscom et al. (2022)[45]*<br><br>**Binary outcomes**: *Jennison and Turnbull (1983)[46], Chang and O'Brien (1986)[47], Duffy and Santner (1987)[48], Emerson (1995)[49], Chang (2004)[50], Jung and Kim (2004)[51], Dallas (2008)[52], Porcher and Desseaux (2012)[53], Kirk and Fay (2014)[54], Yu et al. (2016)[55], Shan (2018)[56], Lloyd (2021, 2022)[35,57], Cao and Jung (2024)[58]* | A (green) | (green) | (green) | (green) | (red) | (amber) |

**Delayed responses/Overrunning**: *Hall and Liu (2002)[59], Hampson and Jennison (2013)[60], Zeng et al. (2015)[61], Shan (2018)[62], Zhao et al. (2015)[63]*

**Repeated measures**: *Lee et al. (2002)[64]*

---

<u>***Repeated confidence intervals***</u>

*Jennison and Turnbull (1984[65], 1989[66], 1999[18])*

- Colours on the right assume that the information levels do not depend on the unknown parameter of interest and that the canonical joint distribution of the group sequential test statistics holds exactly (e.g., for normally distributed data with a known variance - see Jennison and Turnbull, 1999[18])

- If information levels do depend on the unknown parameter of interest, it may not be an interval - see e.g., Jennison and Turnbull (1999)[18]

- RCIs are not tied to any specific stopping rule, so consistency depends on what group sequential test is used to derive the RCI, see e.g., Jennison and Turnbull (1989[66], with discussion). If the test matches the design, then consistency is guaranteed.

- Contains the MLE by construction

**See also**
*Jennison and Turnbull (1990[25], 1991[67]), Davis and Hardy (1992)[68], Fleming and DeMets (1993)[69], Cook (1994)[70], Lee (1995)[71], Hu and Lagakos (1999a, b)[72,73], Posch et al. (2008)[74], Zhao et al. (2009)[75], Zhang et al. (2016)[76], Nowak et al. (2022)[77], Nelson et al. (2022)[15]*

**Time-to-event/Survival outcomes**: *Jennison and Turnbull (1985)[78], Williams (1996)[79], Bernado and Ibrahim (2000)[80]*

**Binary outcomes**: *Lin et al. (1991)[81], Coe and Tamhane (1993)[82]*

**Repeated measures**: *Wei et al. (1990)[83], Jiang (1999)[84]*

**Equivalence tests**: *Jennison and Turnbull (1993)[85]*

---

<u>***Adjusted asymptotic confidence intervals***</u>

*Woodroofe (1992)[19], Todd et al. (1996)[86]*

- Colours on the right are for normally distributed outcomes (with known variance)
- Slight undercoverage can occur for certain parameter values, but conservative coverage is also observed in simulations

- CI is centred around the median unbiased estimator (and not the MLE)

**See also**

**Binary outcomes:** *Todd and Whitehead (1997)[87], Coad and Govindarajulu (2000)[88]*

**Time-to-event/Survival outcomes:** *Coad and Woodroofe (1996[89], 1997[90])*

**Secondary parameters:** *Whitehead et al. (2000)[91]*

---

***Bootstrap/resampling procedures***

*Snapinn (1994)[92], Chuang and Lai (1998[23], 2000[24])*

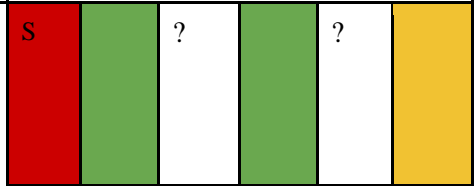

- Can be applied to group sequential trials regardless of stopping boundaries or patient outcome distribution

- Undercoverage can occur and can be substantial

- Computation involves repeated trial simulations, but for calculating a single confidence interval is not time-consuming

- Conditional perspective considered in Pepe et al. (2009)[93], Shimura et al. (2017)[94]

---

**Adaptive group sequential / group sequential with sample size re-estimation**

**Repeated confidence intervals:** *Lehmacher and Wassmer (1999)[95], Wassmer et al. (2001)[96], Wassmer (2003)[97], Hartung and Knapp (2006)[98], Mehta et al. (2007)[99]*

**Exact confidence intervals:** *Wassmer (2006)[100], Brannath et al. (2009)[101], Wang et al. (2010)[102], Gao et al. (2013)[103], Gao and Mehta (2013)[104], Mehta et al. (2019)[105], Gao and Li (2024)[106]*

**See also**
*Hartung and Knapp (2010[107], 2011[108]), and the review article by Nelson et al. (2022)[15]*

- The above references are for adaptive group sequential designs (i.e., group sequential designs additionally encompassing sample size re-estimation), but given the variety of different designs this (sub)class includes, we do not give a traffic light categorisation

| Multi-arm multi-stage designs (with treatment selection) | _Repeated confidence intervals_<br><br>_Zhao et al. (2009)[75], Jaki and Magirr (2013)[109]_<br><br>• Colours on the right assume that the canonical joint distribution of the group sequential test statistics holds (at least approximately)<br><br>• Assumes all promising treatments are taken forward after each interim analysis<br><br>**See also**<br><br>**Single stage multi-arm designs:** _Liu (1995)[110]_ | 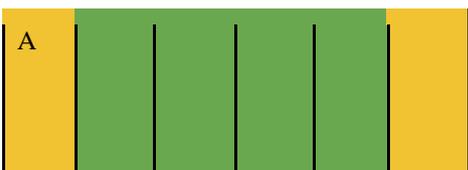 A |
| | _Drop-the-loser designs_<br><br>_Sampson and Sill (2005)[111], Stallard and Todd (2005)[112], Sill and Sampson (2009)[113], Wu et al. (2010)[114], Neal et al. (2011)[115], Magirr et al. (2013)[116], Bowden and Glimm (2014)[117], Kimani et al. (2014)[13], Carreras et al. (2015)[118], Brückner et al. (2017)[119], Whitehead et al. (2020)[120], Gao and Li (2024)[121]_<br><br>• The above references are specifically for drop-the-loser designs, but encompass a variety of different methods for constructing CIs, hence we do not give a traffic light categorisation<br><br>• Some methods e.g., the asymptotic approach of Bowden and Glimm (2014)[117] can have undercoverage<br><br>**See also**<br><br>**Methods using a normal approximation:** _Shun et al. (2007)[122], Bowden and Glimm (2008)[123]_<br><br>**Review article for seamless phase II/III trials:** _Kimani et al. (2014)[13]_ | |
| Response-adaptive randomisation | _Bootstrap/resampling procedures_<br><br>_Rosenberger and Hu (1999)[21], Bandyopadhyay and Biswas (2003)[124], Baldi Antognini et al. (2022)[125], Lane (2022)[126]_<br><br>• Can be applied to multi-arm trials as well as different outcome types<br><br>• Slight undercoverage can occur for some parameter values, but conservative coverage is commonly observed in simulation studies<br><br>• Computation involves repeated trial simulations, but for calculating a single confidence interval is not time-consuming | 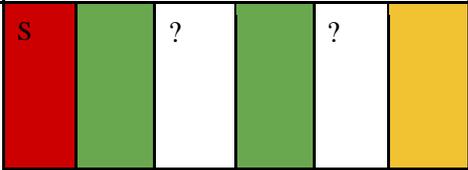 S   ?   ? |

| | | | | | | |
|---|---|---|---|---|---|---|
| | • Both conditional and unconditional bootstrap procedures have been proposed, see Lane (2022)[126] | | | | | |
| | ***Exact confidence intervals***<br><br>*Wei et al. (1990)[127]*<br><br>• Colours on the right are for a two-arm trial with binary outcomes<br><br>• The specific RAR procedure considered in the simulations is the randomised play-the-winner (RPW) rule<br><br>• Undercoverage can occur for large treatment differences, otherwise conservative coverage is observed<br><br>• Computational burden depends on the RAR procedure used | 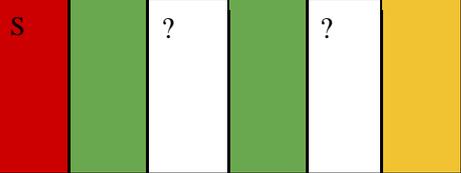 S | | ? | | ? | |
| | ***Adjusted asymptotic confidence intervals***<br><br>*Tolusso and Wang (2011)[128]*<br><br>• Colours on the right are for a two-arm trial with binary outcomes<br><br>• The specific RAR procedure considered in the simulations is the randomised play the winner (RPW) rule<br><br>• Undercoverage can occur for large treatment differences, otherwise conservative coverage is observed<br><br>• Simple closed form expression given for the RPW rule, but in general computational burden depends on the RAR procedure used | 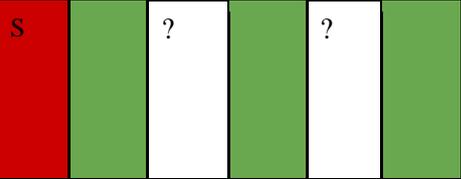 S | | ? | | ? | |
| **Adaptive enrichment designs** | *Brannath et al. (2009)[129], Rosenblum (2013)[130], Wu et al. (2014)[131], Wassmer and Dragalin (2015)[132], Kimani et al. (2020)[133]*<br><br>• The above references encompass a variety of different methods for constructing CIs, hence we do not give a traffic light categorisation<br><br>• These methods in general can be very computationally intensive | | | | | | |

*Table 1: Summary of properties of different methods of constructing CIs, categorised by the broad class of adaptive design.*

Looking at the summary of the literature review as a whole, the number of papers proposing methods for constructing adjusted CIs for ADs has grown quite rapidly in the past 15 or so years. In terms of the properties described by the traffic light system, in terms of coverage it is generally clear (at least by simulation) which methods result in CIs that have correct coverage as well as over- or undercoverage. We caution however that even methods that are 'green' above rely on the assumptions used to derive the CIs holding exactly, and that methods that are 'red' will have differing levels of undercoverage.

All of the main CI methods above were 'green' in terms of being an interval and being informative, but specific subcases (not covered by the traffic light system) can be 'red', as an example of methods possibly returning the empty set (and hence also being uninformative) see Fan and DeMets (2006)[28] and Hartung and Knapp (2010)[107] in the context of conditional CIs for group sequential designs. In the context of RCIs for group sequential designs, Brookmeyer & Crowley (1982)[134] show how in rare cases the RCI may not be an interval if the information levels depend on the parameter of interest $\theta$, with further discussion on this point in Jennison and Turnbull (1999)[18].

In contrast, consistency is harder to assess, with this criterion being unclear for all classes of ADs apart from group sequential designs. This also applies to the criterion of the MLE being included in the CI, reflecting the gap sometimes seen in the literature on point estimation and the literature on CIs for ADs. Finally, all methods (except for the adjusted asymptotic CI for response-adaptive randomisation) were 'yellow' in terms of computation, although this encompasses a wider range of computational complexity.

By far the majority of the proposed methodology for CIs has focused on group sequential designs, which is understandable given their long history and widespread use. Other classes of ADs have received comparatively little attention when it comes to adjusted CIs, which is reflected in the unclear properties for some of the CI methods. Finally, most of the methodology has focused on (at least asymptotically) normally-distributed outcomes or binary outcomes, with comparatively few proposals tailored for trials with time-to-event outcomes.

# 5. Discussion

In our literature review of the methods for constructing CIs for ADs, we found that there is a growing body of work proposing and evaluating a range of CIs for a variety of ADs. Our hope is that this paper, combined with the annotated bibliography given in the Supporting Information, provides an easily accessible and comprehensive resource for trialists and methodologists working on ADs. However, statistical software and code to calculate adjusted CIs unfortunately remains relatively rare (see the annotated bibliography), which is an obstacle to the uptake of methods in practice (see Grayling and Wheeler, 2000[135]). In addition, for more complex or novel ADs, adjusted CIs may not exist in the literature.

From a methodological perspective, while CIs for group sequential designs are very well-developed, this is much less the case for other classes of ADs. In particular, for response-adaptive randomisation and adaptive enrichment designs only a handful of papers proposing adjusted CIs exist. More generally, while coverage is always reported in simulation studies for CIs, the other desirable properties and related performance measures described in Section 4.3 are much less often reported (as also reflected in the traffic light system results in Table 1). Hence, we encourage methodologists working on new CI methods to consider reporting a wider variety of performance measures. In the future it would also be helpful to have methodological proposals around how to appropriately combine different metrics/performance measures of interest.

We note that our focus in this paper has exclusively been on frequentist CIs. We have not discussed the construction of other types of intervals, e.g., prediction intervals and tolerance intervals (see e.g., Vardeman, 1992[136] and Krishnamoorthy & Mathew, 2009[137]) or Bayesian credible intervals (Meeker et al, 2007[138]). Nor have we addressed fixed-width CI construction (i.e., where the design of the trial itself is chosen to achieve a CI with a desired coverage and of a certain width, through a minimal sample size), all of which have had very limited discussion in the context of ADs.

In part II of this paper series, we explore the practical considerations surrounding the use of CIs for ADs. There, we illustrate their application to a two-stage group sequential trial design. We also provide a set of guidelines for best practice, considering the use of CIs in ADs from the design stage through to the final reporting of results.

## Acknowledgements

T Jaki and DS Robertson received funding from the UK Medical Research Council (MC_UU_00002/14 and MC_UU_0040_03). B Choodari-Oskooei was supported by the MRC grant (MC_UU_00004_09 and MC_UU_12023_29). The Centre for Trials Research receives infrastructure funding from Health and Care Research Wales and Cancer Research UK.

## Data Availability Statement

All of the data that support the findings of this study are available within the paper and supporting information. For the purpose of open access, the author has applied a Creative Commons Attribution (CC BY) licence to any Author Accepted Manuscript version arising.